\documentclass{an}
\usepackage{graphicx}
\usepackage{times}
\Volume{00}              
\Year{0000}              
\Month{00}               
\Pagespan{000}{000}      

\begin{document}
\lhead[\thepage]{A.N. S.F.S\'anchez: The Euro3D Visualization Tool}
\rhead[Astron. Nachr./AN~{\bf XXX} (200X) X]{\thepage}
\headnote{Astron. Nachr./AN {\bf 32X} (200X) X, XXX--XXX}

\title{E3D, The Euro3D Visualization Tool I:\\ 
Description of the program and its capabilities
}

\author{S.F. S\'anchez\inst{1}}
\institute{Astrophysikalisches Institut Potsdam, And der Sternwarte 16, 14482
  Potsdam, Germany }
\date{Received {date will be inserted by the editor}; 
accepted {date will be inserted by the editor}} 

\abstract{We present the first version of E3D, the Euro3D visualization tool
for data from integral field spectroscopy. We describe its major characteristics, 
based on the proposed requirements, the current state of the project, and some
planned future upgrades. We show examples of its use and capabilities.
}
\correspondence{ssanchez@aip.de}

\maketitle

\section{Introduction}

The Euro3D Research Training Network (RTN) (\cite{net02}) was put forward with
the intention to promote integral field spectroscopy (IFS), or ``3D''
spectroscopy, and to help making it a common user technique. In order to
accomplish this, one of the major tasks was identified as the need of
providing standard software tools for the visualization and analysis of
datacubes. These tools should be general enough to be entirely independent of
the origin of data, i.e.\ 3D instrument.  Previously, a heterogenous
collection of instrument-specific data formats and software tools (e.g.\ 
XOASIS), proprietary software packages and a lack of any standard have
hampered a break-through of this powerful observing method, leaving it merely
as an expert technique with comparatively limited scientific impact.

Recognizing the importance of this problem, a work plan was devised to start
creating a package of tools for the analysis and visualization of IFS data.
Entitled {\it 3D Visualization}, Task~2.2 of this work plan foresees the
development of a programme, which should be capable of reading, writing, and
visualizing reduced data from 3D spectrographs of any kind. We have named this
tool ``{\bf E3D}''.

Here we present the current status of the project, give a brief description of
the programme as it is now, point out some requirements which have not yet been
met, and explain some problems that were encountered during the development.  
We also present some examples with real data, trying to explore the potential 
of the tool already at its first stage of development.

\section{Background}

One of the major problems for the development of a standard visualization tool
is the lack of a standard data format. Every group has developed its own {\it
  3D data format}, both for the spectral and the position information (cubes,
FITS images, FITS tables, MIDAS images, etc...).  In order to overcome this
problem, the RTN has proposed a unified data format, the ``Euro3D Data
Format'' (\cite{kp03}; \cite{kp03b}).  Taking into account previous experience
from more than a decade of operating 3D instrumentation in the visible and the
near-infrared, this data format is supposed to cover most foreseeable
requirements of existing and future instruments. The Euro3D visualization tool
was written specifically to make use of this data format.

It was the scope of the network from the very begining to provide a freely
distributed software, that could be installed/used on the largest possible
number of computers. This prevents us from developing the software in any
comercial (e.g., IDL) or non comercial environment (e.g., MIDAS) that could
create a long-term dependence or limit its use. The possibility of
using/adapting a previous existing tool (like DS9, XIMAGE or GIPSY) was
considered. However, the expecific requeriments of IFS prevented us from
choosing this solution. A major caveat was the requisite of that tools to
handle with regular gridded data, like datacubes, which force us to
interpolate (i.e., alter) the data to visualize them. Due to all these reasons
it was decided to write a stand-alone software in C.

A C-coded library (``LCL'') was developed to handle the input/output of data
on the proposed format (\cite{fe03}).  This library allows to read and write
not only Euro3D format files, but also reads/writes single spectra,
monochromatic datacube slices, FITS images, and FITS tables.  We have tested
different graphical libraries (NCARG, PLPLOT, X11 low-level routines,etc.) and
created different prototypes based on these various libraries. As a result, it
was decided to use PGPLOT, mainly due to its flexibility, portability, and in
particular its capability to interact with Tcl/Tk. The latter property allowed
us to implement a scripting capability.

\section{Requirements}

Decisions upon the specification of E3D were made after extensive discussion
in various RTN meetings. The following summary explains the main requirements:

\begin{enumerate}
\item  Display all the spectra stored on the file as a single 2D image ({\it
    staked spectra}, one spectrum per row).
  
\item Display different spatial representations of the data ({\it maps}) from
  the stacked spectra representation:

  \begin{enumerate}
  \item Display maps built from a single spectral pixel ({\it monochromatic
      maps}) or from integration over a given spectral range ({\it
      polychromatic maps}), using the geometry of the spatial elements ({\it
 spaxels}: fibers, lenslets...). {\it done}
  \item Display interpolated images {\it done}
  \item Page handling, blinking {\it done}
  \item Display maps with overlay of the spaxel shapes {\it done}
  \item Correct for atmospheric refraction {\it done}
  \item Weighted summation of maps, with rejection of noisy spaxels {\it work in progress}
  \item Load and display regular FITS spectra, images and tables {\it work in progress}
  \item Simple statistical tools {\it work in progress}
  \end{enumerate}

\item Select spectra from map representation:
\begin{enumerate}
  \item  Investigate interaction between windows, events {\it done}
  \item  Window Layout {\it ongoing development}
\end{enumerate}

\item Alternative representations: 
  \begin{enumerate}
  \item  pseudo long-slit {\it done}
  \end{enumerate}

\end{enumerate}

In addition, more demands have emerged in the course of discussion. It was
demanded that E3D were built with a modular phylosophy that allows to integrate
different packages on the future. E3D should be able to interact with the
major astronomical data analysis packages, like IRAF/PyRAF or IDL. A Shared
Memory Server (SHM) was integrated into E3D for this propose, although its
capabilities have not been already fully tested. Additional communication
methods have been developed and tested, based on the scriptable capabilities
of E3D.

\section{Characteristics}

E3D comprises a C-coded core, with three main elements:
\begin{itemize}
\item The {\tt Euro3D.o} library of low-level functions, including the
  routines calling the Euro3D I/O library, the SHM routines, basic functions
  needed for plotting and analizing the data. This library can be invoked to 
  create another program, which needs to read/write Euro3D data, or uses the 
  SHM routines.
\item The {\tt tk\_e3d } Tcl/Tk interpreter. This is a standalone programme
  that creates its own Tcl/Tk interpreter, adding Euro3D routines to the
  standard Tcl ones.  These routines invoke the C-functions included in {\tt
    Euro3D.o} from Tcl, and they can be used for different proposes: e.g.,
  load/save Euro3D format files, plot single or coadded spectra, plot
  monochromatic/polychromatic maps, interpolate these, save maps in FITS
  format.
\item A number of stand-alone C-coded tools that help to handle the
  Euro3D format.  Perhaps the most interesting routine is {\tt any2Euro3D},
  transforming single Group IFS data from any instrument into the Euro3D Format.
\end{itemize}

\begin{figure}
\resizebox{\hsize}{!}
{\includegraphics[width=7cm,angle=90]{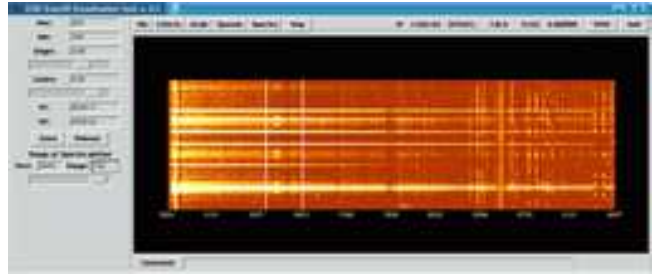}}
\caption{Stacked Spectra Inspector. 
This is the main GUI window, from which all the others can be created. It
displays all the spectra contained in the 3D file stacked row-by-row.}
\label{raw_ins}
\end{figure}

\begin{figure}
\resizebox{\hsize}{!}
{\includegraphics[width=7cm,angle=90]{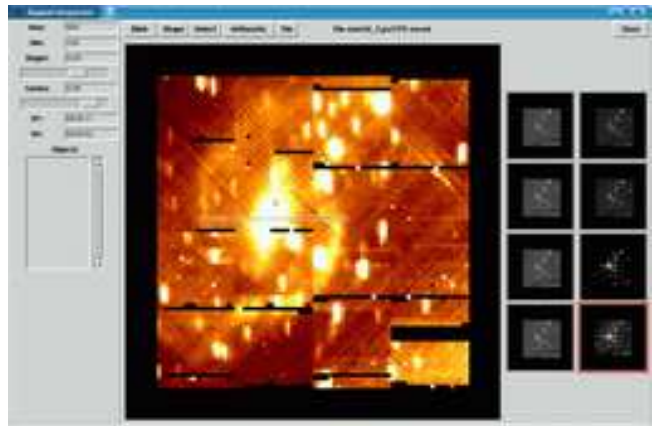}}
\caption{Spaxels Inspector. This is the GUI for plotting monochromatic or polychromatic 
maps. It is possible to select different spaxels, to be displayed subsequently on the 
Spectral Inspector.}
\label{spax_ins}
\end{figure}

\begin{figure}
\resizebox{\hsize}{!}
{\includegraphics[width=7cm,angle=90]{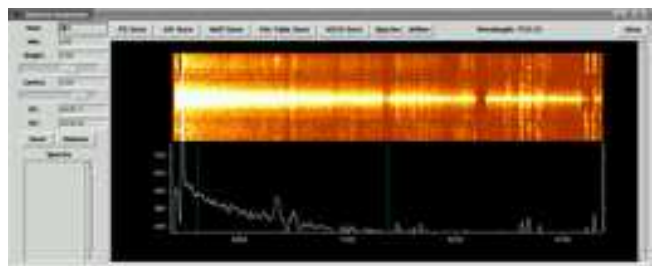}}
\caption{Spectral Inspector. This is the GUI for plotting spectra.}
\label{spec_ins}
\end{figure}

Together with the C-core, we have coded a Tcl/Tk Graphical User
Interface (GUI), that uses the Euro3D-Tcl routines ({\tt tk\_e3d.tcl}). The GUI
comprises three main windows:

\begin{itemize}
\item The Main window or Stacked Spectra Inspector. It comprises the main
  Menu with different options to handle Input/Output and different
  representations of the data. The data are automatically plotted in its
  canvas on the Stacked Spectra representation once loaded. A command-line
  prompt has been included in this window to directly call Euro3D-Tcl
  routines. Figure \ref{raw_ins} shows an image of this window.
\item The Spaxels Inspector. It comprises a main canvas for plotting
  monochromatic/polychromatic datacube slices, which have been selected previously
  from the main window or from the Spectral Inspector (see below).  
  It includes 8 minor canvas for buffering the last created maps. It also
  includes a menu with different options to handle spaxels/maps,
  including the spaxel selection, object creation and the selection of
  different methods to show the data. Figure \ref{spax_ins} shows a screen-shot of
  this window.
\item The Spectral Inspector. It comprises a main canvas for plotting
  the spectra corresponding to spaxels which were selected from any of the two 
  previous windows. It also includes a menu with different options to handle the 
  spectra.
  Figure \ref{spec_ins} shows an image of  this window.
\end{itemize}

As explained above the GUI has been coded in Tcl/Tk using the Euro3D-Tcl
routines. This is probably the most powerful outcome of the adopted
programming philosophy, since a scripting capability for handling Euro3D data
is included ``for free''. This means that any user can create his/her own
Tcl-coded scripts by making use of the Euro3D-Tcl routines, and run them 
by invoking {\tt tk\_e3d }. 

\subsection{The Euro3D-Tcl routines: An example}

It is beyond the scope of this article to explain in detail all of the existing
Euro3D-Tcl routines. However, as an instructive example, we show a script
which creates a {\it movie} or a {\it sequence} of monochromatic slices,
running through the Euro3D datacube:

{\small\tt

01 \#!{PATH}/user/bin/tk\_e3d

02 create\_env euro3d "/XSERV"

03 euro3d load\_file {E3D\_FITSFILE}

04 set data [euro3d ask\_e3d\_info]

05 set start\_w [lindex \$data 2]

06 set delta\_w [lindex \$data 4]

07 set i\_1 0

08 set i\_2 1000

09 euro3d plot\_spaxels -1 25 1 0 10 heat 0.65 0.60 -1 1 1 4 0

10 for \{set x \$i\_1\} \{\$x<\$i\_2\} \{incr x 10\} \{

11 set x2 [expr \$x+10]

12 set w [expr ((\$x+1.5)*\$delta\_w+\$start\_w)]

13 euro3d plot\_spaxels -1 35 1 \$x \$x2 heat 0.65 0.7 1 1 1 4 1  

14 \}

15 exit

}

Line 1 calls the interpreter, {\tt tk\_e3d}. Lines 2 and 3 create the Euro3D
environment, with a permanent XWIN, and load the Euro3D file. Line 4 loads the
descriptor of the file (e.g.\ wavelength range, number of spaxels, etc.) from
the list {\tt data}. Line 5-8 define some variables for later use (e.g.\ 
starting wavelength, {\tt start\_w}, wavelength step, {\tt delta\_w}, and the
first and last spectral pixel to plot, {\tt i\_1} and {\tt i\_2}).  Line 9
plots the first averaged polychromatic map (mean taken over a 10 pixel-wide
window). Lines 10-14 corresponds to a loop over the spectral pixels (from {\tt
  i\_1} to {\tt i\_2}), with a step of 10 pixels. At each step, the
corresponding, averaged polychromatic map is displayed.

\begin{figure}
\resizebox{\hsize}{!}
{\includegraphics[width=1.0\textwidth]{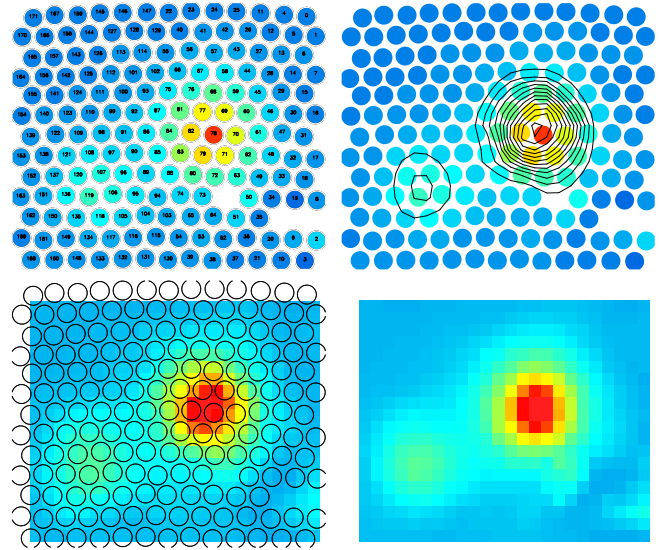}}
\caption{{\bf Top-Left:} Polychromatic map INTEGRAL data of HES 1104-185, using the spaxel
  representation. {\bf Top-Right:} Same map including a countour plot of the
  data. {\bf Bottom-Left:} Interpolated representation of the same map,
  using a Spline interpolation routine. The original spaxels pattern is
  overplotted.  {\bf Bottom-Right:} Similar interpolation, without the spaxels
  pattern, and using a Natural Neighbour interpolation routine.}
\label{example}
\end{figure}

\begin{figure}
\resizebox{\hsize}{!}
{\includegraphics[width=6cm,angle=-90]{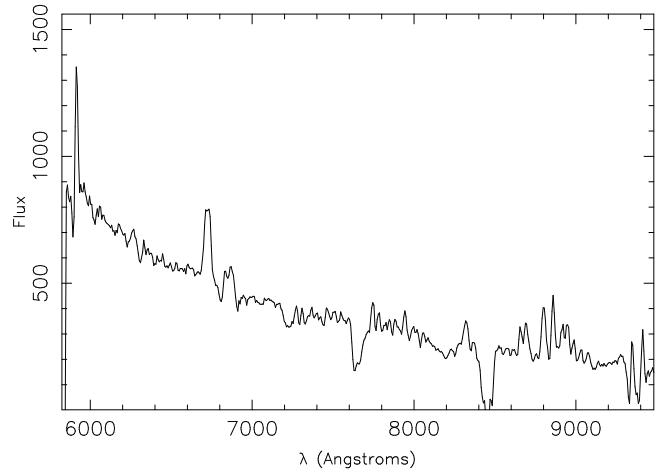}}
\caption{Spectrum of the central square region of
  1.5$\arcsec$$\times$1.5$\arcsec$ of IRAS 13031-5717, observed with
  VIMOS-IFU. The flux is in an arbitrary scale. }
\label{spectrum}
\end{figure}

\subsection{E3D by examples}

E3D actually performs relatively simple routines, like spaxel selection or
image reconstruction. Figure \ref{example} shows an example of four different
representation of the maps. The top-left panel shows a spaxel representation
of a slice cut of INTEGRAL data taken on the gravitational lens HES 1104-185
(\cite{go03}).  Each spaxel has a radius of 0.27$\arcsec$. The top-right
pannel shows the same representation including a contour plot. For displaying
the contour plot it is needed to interpolate the data, creating a regular
gridded map. Five different interpolation algorithms are already available on
E3D. The two bottom pannels of Fig.\ref{example} show two examples of these
interpolation algorithms, both using a 0.3$\arcsec$/pixel grid.

We have included different representations of the spectra, both in a pseudo
slit-spectra form (a spatial cut in the datacube, mimicking a slit-spectrum
observation), as seen in Fig. \ref{spec_ins}, or as standard spectrum plot,
Fig. \ref{spectrum}. These two figures show the averaged spectra of the
central region of IRAS 13031-5717, an interacting galaxy at $z\sim$0.02
(\cite{san03}). This object was observed with VIMOS at the VLT, using the
low-resolution red setup. We have reduced the data with a modified version of
the P3d data-reduction software (\cite{be01}).  These data have been used as
examples in Figures \ref{raw_ins}, \ref{spax_ins} and \ref{spec_ins}.

\begin{figure}
\resizebox{\hsize}{!}
{\includegraphics[width=6.8cm,angle=-90]{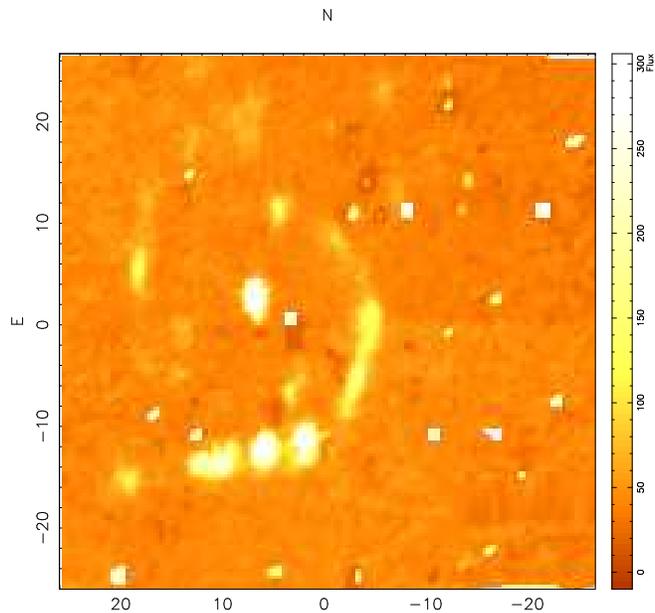}}
\caption{Interpolated H$\alpha$ map of IRAS 13031-5717 observed with
  VIMOS. The map has been created by subtracting a polychromatic map the
  continuum near H$\alpha$ from a map centered on the emission line. The
  interpolation grid consists of pixels of 0.3$\arcsec$ size. }
\label{arith}
\end{figure}

A number of simple analysis tools has been added to E3D. Among these tools are
the {\tt specarith} and {\tt spaxarith } routines. Both routines allow one to
perform arithmetic operations between selected spectra and selected
polychromatic maps, respectively. Figure \ref{arith} shows an example of {\tt
  spaxarith}. This figure shows the H$\alpha$ map of IRAS 13031-5717. To
create this map we have subtracted a continuum map near H$\alpha$ from a map
centered on the emission line. This option is now included in the GUI.

\section{Future work}

We have designed E3D to be a data visualization and data analysis tool. It is
our goal to integrate as many different tasks of the Euro3D software package
as possible into E3D. In the end, this strategy will provide a powerful
analysis/visualization tool. There are a number of bugs still to be fixed,
some of which have been identified. A few basic requirements are still on the
queue.  For example, it has still to be decided how to handle different
wavelength units (for now: Angstroms), and different spatial units (for now:
arcsec).  There is a need for improved zooming capabilities.  We have to think
how to treat the data quality flags, and which are the best and most flexible
defaults.  Different methods of selecting spaxels (area selection) have been
proposed, but they have not been coded.  It is still under discussion how to
handle science tables, and how to plot their contents.

So far, we have tested E3D with data from a variety of instruments :
INTEGRAL (\cite{ar98}), OASIS, PMAS (\cite{ro00}), SAURON, SparsePak
(\cite{ber03}), SPIFFI, TIGER, VIMOS, and with different mosaic patterns (e.g.
\cite{san03}). Some memory bugs and overloading problems have been detected,
rendering the program not very efficient for massive reloads of big frames.
We need further investigations of how to interact with external packages (IDL,
PyTHON, ...), and further tests with the SHM are needed.

However, given this early stage of development, E3D seems to be a promising
tool, which has already proven to be useful for visualize and help the
analysis of real 3D data.

\acknowledgements

This project is part of the Euro3D RTN on IFS, funded by the European
Commission under contract No.  HPRN-CT-2002-00305. I like to acknowledge: S.
Foucaud for his help and M.  Horrobin for his tests.  A.P\'econtal-Rousset, P.
Ferruit and the entire Lyon group for their advice, help, and their marvelous
work on the LCL library. M.M.Roth for his help on editing this manuscript.

\begin{thebibliography}{}
\bibitem[Arribas et al. 1998]{ar98} Arribas S., Carter, D., Cavaller, L., et
  al.: 1998, Proc. SPIE, 3355, 821
\bibitem[Becker 2001]{be01} Becker, T.: 2001, PhD Thesis, Astrophysikalisches Institut Potsdam, Germany
\bibitem[Bershady et al. 2003]{ber03} Bershady, M.A., Andersen D.R., Harker,
  J., Ramsey, L.W., Verheijen, M.A.W., 2003, PASP, submitted 
\bibitem[Pecontal-Rousset et al. 2003]{fe03} Pecontal-Rousset, A., Ferruit, P., et al., 2003, Euro3D Science Workshop, 21-23 May 2003, IoA, Cambridge, AN, these proceedings.
\bibitem[G\'omez et al. 2003]{go03} G\'omez, P., Mediavilla, E.,  S\'anchez,
  S.F., et al., 2003,  Euro3D Science Workshop, 21-23 May 2003, IoA,
  Cambridge, AN, these proceedings.
\bibitem[Kissler-Patig et al. 2003a]{kp03} Kissler-Patig, M., Copin, Y., Ferruit, P., P\'econtal-Rousset, A., Roth, M.M., 2003, Euro3D Data Format Definition, Euro3D Documentation.
\bibitem[Kissler-Patig et al. 2003b]{kp03b} Kissler-Patig, M., Copin, Y.,
  Ferruit, P., P\'econtal-Rousset, A., Roth M.M.,  2003,
  Euro3D Science Workshop, 21-23 May 2003, IoA, Cambridge, AN, these proceedings.
\bibitem[Roth et al. 2000]{ro00} Roth, M.M., Bauer, S., Dionies, F., et al. : 2000, in Proc. SPIE, Vol. 4008, 277-288
\bibitem[S\'anchez et al. 2003]{san03} S\'anchez, S.F., Christensen, L.,
  Becker, T., Kelz, A., Jahnke, K., Benn, C.R., Garc\'\i a-Lorenzo, B., Roth,
  M.M. : 2003, AN, these proceedings.
\bibitem[Walsh \& Roth 2002]{net02} Walsh, J.~R.~\& Roth, M.~M.: 2002, The Messenger, 109, 54 
\end{thebibliography}
\end{document}